%% file: KWT_ISIT_final_rev.tex
\documentclass[conference]{IEEEtran}
\ifCLASSINFOpdf
\else
\fi
\usepackage[dvips,dvipdfm]{graphicx}
\usepackage{color}
\include{commands}

\begin{document}
\title{Statistical Mechanical Analysis of a Typical Reconstruction 
Limit of Compressed Sensing}

\author{\IEEEauthorblockN{Yoshiyuki Kabashima}
\IEEEauthorblockA{Department of Computational Intelligence \\and 
Systems Science\\
Tokyo Institute of Technology\\
Yokohama 226--8502, Japan\\
Email: kaba@dis.titech.ac.jp}
\and
\IEEEauthorblockN{Tadashi Wadayama}
\IEEEauthorblockA{Department of Computer Science \\
Nagoya Institute of Technology \\
Nagoya 466--8555, Japan \\
Email: wadayama@nitech.ac.jp}
\and
\IEEEauthorblockN{Toshiyuki Tanaka}
\IEEEauthorblockA{
Department of Systems Science \\
Kyoto University\\
Kyoto 606--8501, Japan\\
Email: tt@i.kyoto-u.ac.jp}}


%Telephone: (800) 555--1212, Fax: (888) 555--1212}


\maketitle

\begin{abstract}
%\boldmath
We use the replica method of statistical mechanics to examine 
a typical performance of correctly reconstructing $N$-dimensional sparse vector $\bx=(x_i)$
from its linear transformation $\by=\bF \bx$ of $P$ dimensions  
on the basis of minimization of the $L_p$-norm $||\bx||_p=
\lim_{\epsilon \to +0} \sum_{i=1}^N |x_i|^{p+\epsilon}$. 
We characterize the reconstruction performance by the critical relation of the successful reconstruction 
between the ratio $\alpha=P/N$ and the density $\rho$ of non-zero elements in $\bx$ 
in the limit $P,\,N \to \infty$ while keeping $\alpha \sim O(1)$ and 
allowing asymptotically negligible reconstruction errors. 
%When $\bF$ is composed of independently and identically distributed 
%random variables of zero mean and a fixed variance, 
%the critical relation $\alpha_c(\rho)$ for $p=1$ accords with 
%that obtained by Donoho 
%({\em Discrete and Computational Geometry}, vol.~35, pp.~617--652, 2006)
%who utilized a technique of 
%combinatorial geometry and allowed no reconstruction errors.  
{
We %also 
show that the critical relation $\alpha_c(\rho)$
% does not change 
%
holds universally
as long as 
$\bF^{\rm T}\bF$ can be characterized asymptotically by a rotationally invariant random matrix ensemble
and $\bF \bF^{\rm T}$ is typically of full rank. 
This supports the universality of the critical relation observed by Donoho and Tanner 
({\em Phil. Trans. R. Soc. A}, vol.~367, pp.~4273--4293, 2009; 
arXiv: 0807.3590)
for various ensembles of compression matrices. 
}
\end{abstract}


% \fi
\IEEEpeerreviewmaketitle

\section{Introduction}
Let us suppose a situation that an $N$-dimensional  vector $\bx^0 \in \mR^N$
is linearly transformed into a $P$-dimensional vector $\by \in \mR^P$, where 
$\by=\bF \bx$, by using a $P \times N$ matrix $\bF \in \mR^{P \times N}$. 
When $\alpha =P/N < 1$, it is generally impossible to reconstruct the original vector 
$\bx^0$ from $\by$ and $\bF$ correctly. However, if $\bx^0$ is sparse, 
which means that the fraction $\rho$ of non-zero elements in $\bx^0$ is 
less than unity, a correct reconstruction may be possible by utilizing 
prior knowledge of the sparsity. 
The framework for making efficient use of such a possibility is often termed
{\em compressed (compressive) sensing} \cite{CS}. The research history of this scheme 
is rather long and dates back to the 1970s \cite{Claerbout,Santosa}. 
However, since the publication of a series of influential papers in the middle of 
the 2000s, the horizon of this field has rapidly expanded 
\cite{Donoho1,Donoho2,CandesTao3,CandesTao1,CandesTao2}. 

For exploiting the sparsity in reconstructing $\bx^0$ from $\by$ given $\bF$, 
reconstruction schemes based on minimization of the so-called $L_p$-norm of $\bx=(x_i) \in \mR^N$
\begin{eqnarray}
&&||\bx||_p=\lim_{\epsilon \to +0} \sum_{i=1}^N |x_i|^{p+\epsilon}\cr
&&= \! \left \{  \!
\begin{array}{ll}
\sum_{i=1}^N |x_i|^p, &  \! p > 0 \cr
\mbox{the number of nonzero elements in $\bx$}, &  \! p=0 
\end{array}
\right .
\label{Lp_norm}
\end{eqnarray}
under a linear constraint
$
\bF \bx=\by
$
have been actively explored.
We will hereafter refer to these schemes as {\em $L_p$-reconstruction}.  
In particular, much attention has been paid to $L_1$-reconstruction 
since  (\ref{Lp_norm}) under the linear constraint 
can be minimized by using convex optimization methods with a computational 
cost that is polynomial in $N$ for $p=1$ \cite{Danzig,HauptNowak,Lasso,ChenDonohoSaunders}. 

The following result on $L_1$-reconstruction was reported recenly 
\cite{CandesTao3,CandesTao2,Candes2008}. 
Let us suppose that the number of non-zero elements in $\bx^0$ is upper-bounded by $S$. 
Let us also assume that $\bF$ is composed of independently and identically distributed (i.i.d.) 
Gaussian random variables of zero mean and a fixed variance. 
In addition, we shall consider that the reconstruction is successful if and only if
the reconstructed vector
$\widehat{\bx}$ exactly accords to $\bx^0$. 
Accordingly, if the two inequalities
\begin{eqnarray}
&&\frac{2S}N\ln \left (\frac{N}{2S} \right )+\frac{2S}{N}+\frac{1}{N}
\ln (2S)\cr
&&-\frac{P}{2N}\left 
(2^{1/4}-1-\sqrt{\frac{2S}{P}}\right )^2 < 0 
\label{sufficient1}
\end{eqnarray}
and
\begin{eqnarray}
2^{1/4}-1-\sqrt{\frac{2S}{P}} > 0 
\label{sufficient2}
\end{eqnarray}
hold simultaneously, the probability of the $L_1$-reconstruction
failing vanishes as $N$ tends to infinity. 
The previous studies \cite{Donoho2006,DonohoTanner2009}
obtained the critical relations between $\alpha$ and $\rho$ 
for typical and worst cases 
when $\bx^0$ is randomly generated 
under the same assumption of $\bF$ and for the same criterion on the successful reconstruction 
as $P,\,N \to \infty$ while maintaining $\alpha \sim O(1)$. 

The above two results have one thing in common in the sense that 
the critical relations are evaluated under
a success criterion that allows no reconstruction errors. 
However, in performance evaluations of large systems, 
it may be more plausible to permit asymptotically negligible 
errors in characterizing the success of reconstruction. 
In addition, it may be natural to ask how good a reconstruction
can be obtained by using the schemes of $p \ne 1$
and/or other choices of compression matrix $\bF$. 

The purpose of this paper is to answer these questions.
More specifically, we utilize the replica method of 
statistical mechanics to explore the typical performance 
of the $L_p$-reconstruction while allowing asymptotically negligible 
reconstruction errors in the large system limit
for a class of compression matrices \cite{KWT}.  

This paper is organized as follows. The next section
introduces the model that we will examine. 
Section III outlines the analysis.  
In section IV,  the results of the analysis are shown in conjunction with an 
experimental validation. The final section is devoted to a summary.

\section{Model definition}
We will consider the following simple scenario of 
compressed sensing. Each component of the original vector
$\bx^0$, $x_i^0$, is i.i.d. following a distribution
\begin{eqnarray}
P(x)=(1-\rho) \delta(x)+\rho r(x) 
\label{prior}
\end{eqnarray}
where $\delta(x)$ denotes Dirac's $\delta$ function and 
$r(x)$ is an arbitrary distribution with finite first- and second-order 
moments and without a finite mass at the origin. For simplicity, we assume that the second moment 
about the origin is unity without loss of generality.
% In order to take more general possibilities into account
% than in the earlier studies \cite{CandesTao3,CandesTao2,Candes2008, Donoho2006,DonohoTanner2009}, 
%
{
Motivated by the observed universality of the critical relation 
between $\alpha$ and $\rho$ for the successful reconstruction, 
which was reported by earlier studies
\cite{DonohoTannerPhilTrans2009,DonohoTannerDCG2008}
for various compression matrices,
we %assume that the compression matrix $\bF$ is characterized by 
characterize $\bF$ as
\begin{eqnarray}
\bF=\bU \bD \bV^{\rm T}
\label{SVD}
\end{eqnarray}
utilizing a form of singular value decomposition %\cite{ShinzatoKabashima2008,ShinzatoKabashima2009}, 
\cite{ShinzatoKabashima2008}, 
where $\bU$ and $\bV$ are samples of uniform distributions over 
$P \times P$ and $N \times N$ orthogonal matrices, respectively, 
which are independent of each other and of $\bD$.  
}
${\rm T}$ denotes the matrix transpose operation. 
$\bD$ is a $P \times N$ diagonal matrix 
such that the eigenvalues of the $P \times P$ matrix $\bD\bD^{\rm T}$
(or $\bF \bF^{\rm T}$) asymptotically 
follow a certain distribution $f(\lambda)$, the support of 
which is defined over a bounded interval in $\lambda >0$ as $P=\alpha N$ 
tends to infinity. 
The situation in which $\bF$ is composed of i.i.d. random variables
of zero mean and a fixed variance corresponds to 
the case that $f(\lambda)$ is provided by a distribution 
of the Marchenko-Pastur type \cite{MarchenkoPastur}. 
Let us suppose that data $\bbm^0$ (which may represent 
image, sound, etc.) is expressed as $\bbm^0=\bV^{\rm T} \bx^0$ by
utilizing an orthogonal basis $\bV^{\rm T}$ and a sparse coefficient 
vector $\bx^0$. 
The assumption of (\ref{SVD}) corresponds to 
a situation in which $\bx^0$ is inferred from projections of $\bbm^0$ to randomly 
chosen $P$ orthogonal directions specified by $\bU$ in conjunction with certain signal 
amplification expressed by $\bD$. 

In the following, we will examine the critical relation between $\alpha$ and $\rho$
for correctly reconstructing $\bx^0$ from its compressed expression $\by=\bF \bx^0$
by the $L_p$-reconstruction, in particular, for cases of $p=0, 1$ and $2$. 
However, such a critical relation disappears when $\by$ is smeared by 
noise of non-negligible strength \cite{Rangan}.
% 
%which is examined utilizing the replica method in \cite{Rangan}.
%

\section{Outline of Replica Analysis}
A posterior distribution 
\begin{eqnarray}
P_{\beta}(\bx|\by)=\frac{e^{-\beta ||\bx||_p} 
\delta\left (\bF\bx-\by \right )}{Z(\beta;\by)} 
\label{boltzmann}
\end{eqnarray}
given a compressed expression $\by$ and compression matrix $\bF$
is the basis of our assessment. 
The normalization factor $Z(\beta;\by)=\int e^{-\beta||\bx||_p}
\delta(\bF \bx-\by)\,d \bx$ acts as the partition function in 
statistical mechanics. 
In the limit $\beta \to \infty$, the posterior distribution (\ref{boltzmann})  
converges to a uniform distribution over solutions of 
the $L_p$-reconstruction which minimize $||\bx||_p$ under 
the constraint $\bF \bx=\by$. 
Therefore, one can explore the performance of the $L_p$-reconstruction 
by examining the properties of the posterior distribution (\ref{boltzmann}) in the limit $\beta \to \infty$. 

A distinctive feature of the posterior distribution (\ref{boltzmann}) is that it depends on predetermined
(quenched) randomness $\by(=\bF\bx^0)$ and $\bF$, which naturally 
leads us to employ the replica method \cite{Dotzenko}. 
More precisely, we evaluate the generating function of the partition function 
$\Phi(n,\beta)=N^{-1}\ln \left [Z^n(\beta;\by)\right ]$ ($n \in \mR$) by analytically continuing 
the functional expressions obtained for $n =1,2,\ldots \in \mN$ to $n \in \mR$, 
where $\left [ \cdots \right ]$ denotes averaging with 
respect to the predetermined random variables $\by$ and $\bF$ (or $\bx_0$, $\bU$ and $\bV$). 
Once the expression of $\Phi(n,\beta)$ is obtained, typical properties of the $L_p$-reconstruction 
can be examined by evaluating the minimized $L_p$-norm (per element), $C_p$, 
with the identity $C_p=-\lim_{\beta \to \infty} \lim_{n \to 0}
\beta^{-1}(\partial/\partial n) \Phi(n,\beta)$. 
{
Similar schemes have been employed for analyzing
various problems concerning information and communication 
\cite{KabashimaSaad,Tanaka2002} including that of noisy sensing \cite{Rangan}.
}

For $n \in \mN$, expanding the $n$-th power of the integral of the partition function yields 
\begin{eqnarray}
\left [Z^n(\beta;\bbs) \right ] = 
\int  
\prod_{a=1}^n \! e^{-\beta ||\bx^a||_p}
\prod_{a=1}^n \delta(\bF\bx^a-\by)
\prod_{a=1}^n d\bx^a. 
\label{nth_replica}
\end{eqnarray}
Employment of the techniques developed in \cite{TakedaUdaKabashima,TakedaHatabuKabashima}, in conjunction with
a heuristic identity $\delta(x)=\lim_{\tau \to +0}
(2 \pi \tau)^{-1/2} \exp (-x^2/(2\tau))$, leads to 
\begin{eqnarray}
&&\lim_{N\to\infty}\frac{1}{N} \ln \left [\prod_{a=1}^n \delta(\bF\bx^a -\by) \right ]_{\bU,\bV} \cr
&& = \lim_{\tau \to +0} \left \{-\frac{n \alpha}{2}\ln (2 \pi \tau) 
+ {\rm Tr} \ G\left (-\frac{{\cal T}(n)}{\tau} \right )  \right \}. 
\label{replica_average}
\end{eqnarray}
Here, $\left [\cdots \right ]_{\bU,\bV}$ denotes an
average with respect to $\bU$ and $\bV$, and 
$G(x)$ is defined on the basis of the Shannon transform of the eigenvalue distribution $f(\lambda)$, as 
\begin{eqnarray}
G(-x)&=&\mathop{\rm extr}_{\Lambda}
\left \{-\frac{\alpha}{2} \int f(\lambda)\ln (\Lambda+\lambda)\,d\lambda+\frac{\Lambda x}{2} 
\right \}\cr
&\phantom{=}& -\frac{1}{2} \ln x -\frac{1}{2}, 
\end{eqnarray}
where $\mathop{\rm extr}_{X}(\cdots)$ means extremization with respect to $X$. 
For $x \gg 1$, this function asymptotically behaves as
$G(-x)\simeq -(\alpha/2) \ln x 
+ \hbox{const.}$ irrespectively of $f(\lambda)$
as long as the support of $f(\lambda)$ is defined over a bounded interval in $\lambda >0$. 
${\cal T}(n)$ is an $n \times n$ matrix whose elements 
are $T_{ab}=N^{-1}(\bx^a-\bx^0) \cdot (\bx^b-\bx^0)$
for $a,b=1,2,\ldots,n$. Under the replica symmetric (RS) ansatz
$N^{-1}|\bx^a|^2=Q$, $N^{-1}\bx^a \cdot \bx^b=q$ and $N^{-1}\bx^0 \cdot \bx^a=m$
for $a (\ne b)=1,2, \ldots, n$, ${\cal T}(n)$ has eigenvalues 
$Q-q-n(Q-2m +Q^0)$ and $Q-q$ of a single degeneracy and $n-1$ denegeracies, respectively,  
where $Q^0=N^{-1}|\bx^0|^2$.  
These make it possible to define the right-hand side of (\ref{replica_average}) for 
$n \in \mR$. 
Combining this and the expression for the volume of variables $\{\bx^a\}$ that satisfy the RS ansatz, 
which can be assessed by using the saddle-point method 
and taking an average with respect to $\bx^0$, finally yields 
\begin{eqnarray}
C_p&=&\mathop{\rm extr}_{\Theta}
\left \{
\frac{\alpha(Q-2m+\rho)}{2 \chi}+\widehat{m}m-\frac{\widehat{Q}Q}{2}
+\frac{\widehat{\chi}\chi}{2} \right .
\cr
&\phantom{=}&
+(1-\rho)\int \phi_p\left (\sqrt{\widehat{\chi}} z;\widehat{Q} \right)\,Dz  \cr
&\phantom{=}&
\left .
+\rho\int \phi_p
\left (\sqrt{\widehat{\chi}+\widehat{m}^2}z;\widehat{Q} \right )\,Dz
\right \}
\label{freeenergy}
\end{eqnarray}
where $\Theta=\{Q,\chi,m,\widehat{Q},\widehat{\chi},\widehat{m}\}$, 
$\chi=\lim_{\beta \to \infty}\beta(Q-q)$, 
$Dz=\exp (-z^2/2 )\,dz/\sqrt{2 \pi}$, and
\begin{eqnarray}
\phi_p(h;\widehat{Q})
=\lim_{\epsilon \to +0}
\left \{\mathop{\rm min}_{x}\left \{
\frac{\widehat{Q}}{2}x^2-h x+|x|^{p+\epsilon}
\right \} \right \}. 
\label{phi_p}
\end{eqnarray}
We have used the fact $Q^0 \to \rho \int r(x^0) (x^0)^2\, dx^0=\rho$, which follows from our assumption
(\ref{prior}) on the distribution $P(x)$. 

\begin{figure}[t]
\setlength\unitlength{1mm}
\begin{picture}(80,30)(0,0)
\put(5,0){\includegraphics[width=10cm]{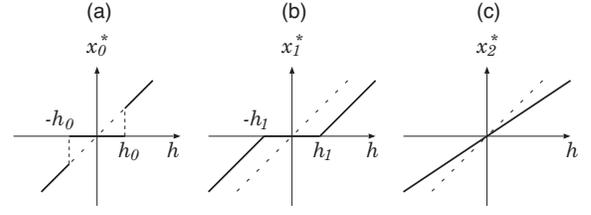}}
\end{picture}
\caption{$x_p^*(h;\widehat{Q})$
for $p=0,1$ and $2$. 
(a): $x_0^*(h;\widehat{Q})=h/\widehat{Q}$ for $|h| > h_0$ and 
$0$, otherwise, where $h_0=\sqrt{2 \widehat{Q}}$. 
(b): $x_1^*(h;\widehat{Q})=(h-h/|h|)/\widehat{Q}$ for  $|h| > h_1$ 
and $0$,  otherwise where $h_1=1$. 
(c): $x_2^*(h;\widehat{Q})=h/(\widehat{Q}+2)$. 
}
\label{fig1}
\end{figure}

At this point, we should note the following issues. 
%%%
{
Firstly, (\ref{freeenergy}) depends on no properties of $\bD$ 
except for the compression rate $\alpha$. 
This means that the asymptotic performance of the $L_p$-reconstruction is 
unchanged as long as $\bF^{\rm T} \bF$ can be asymptotically characterized by 
a rotationally invariant matrix ensemble and $\bF \bF^{\rm T}$ is typically 
of full rank. This is in accordance with the universality on the reconstruction thresholds 
for various matrix ensembles observed in \cite{DonohoTannerPhilTrans2009,DonohoTannerDCG2008}. 
Such universality is, however, limited to the noiseless cases; the reconstruction 
performance does depend on $\bD$ when non-negligible noises exist~\cite{Rangan}.
}
Secondly, $Q$ and $m$ %obtained by the extremization problem 
determined by 
%of 
(\ref{freeenergy}) represent values of 
$N^{-1}\left [|\widehat{\bx}|^2 \right ]$ and $N^{-1} 
\left [\bx^0 \cdot \widehat{\bx} \right ]$, 
respectively, as long as the solution of the $L_p$-reconstruction 
$\widehat{\bx}$ is uniquely determined in typical cases. 
This means that the typical value of the mean square error (MSE) per 
element 
between the original and reconstructed vectors, 
$\mathcal{E}=N^{-1}\left [ |\widehat{\bx}-\bx^0|^2 \right ]$, 
can be assessed as
\begin{eqnarray}
\mathcal{E}=Q-2m+\rho
\label{MSE}
\end{eqnarray}
by using the extremum solution of (\ref{freeenergy}). 
When $\widehat{\bx}$ typically accords with $\bx^0$, 
$Q=m=\rho$ holds, yielding $\mathcal{E}=0$. 
Therefore,  the critical condition between $\alpha$ and $\rho$
for the $L_p$-reconstruction, which can be expressed as 
the critical rate of the reconstruction limit $\alpha_c(\rho)$, can be obtained by 
examining the stability of the successful solution 
$Q=m=\rho$ of (\ref{freeenergy}). 
%
%Secondly, although the objective system is characterized by the asymptotic 
%eigenvalue distribution $f(\lambda)$ of $\bF\bF^{\rm T}$ and the 
%distribution $r(x)$ of non-zero elements in $\bx^0$, 
%(\ref{freeenergy}) depends on neither of them. 
%This implies that the reconstruction limit is  
%invariant in a relatively wide class of systems. 
%
Thirdly, asymptotically negligible MSEs 
are allowed for successful reconstruction in the current analysis, 
whereas no errors are permitted in typical case 
analyses presented in earlier studies \cite{Donoho2006,DonohoTanner2009}. 
This means that our assessment is of utility for examining
how sensitive the performance is to the criterion of the success of 
reconstruction.  
Finally, we must keep in mind that the RS ansatz that is adopted in the 
current analysis is not necessarily guaranteed to be correct. 
For instance, the local stability of the RS saddle point 
is lost against perturbations that 
break the replica symmetry if 
\begin{eqnarray}
&& \frac{\alpha}{\chi^2}
\left ((1-\rho)
\int \left (\frac{\partial x_p^*(\sqrt{\widehat{\chi}}z;\widehat{Q})}{
\partial (\sqrt{\widehat{\chi}}z)}
\right )^2\, Dz \right . \cr
&&\phantom{aa}
\left . 
+ \rho 
\int \left (\frac{\partial x_p^*(\sqrt{\widehat{\chi}+\widehat{m}^2}z;\widehat{Q})}{
\partial (\sqrt{\widehat{\chi}+\widehat{m}^2}z)}
\right )^2\, Dz \right ) > 1
\label{AT}
\end{eqnarray}
holds \cite{AT}. Here, $x_p^*(h;\widehat{Q})=-(\partial/\partial h)
\phi_p(h;\widehat{Q})$, the profile of which is 
shown in Fig.~\ref{fig1} for $p=0,1$ and $2$. 
When the condition (\ref{AT}) holds for the extremum solution 
of (\ref{freeenergy}), the RS treatment is not valid 
and one has to explore more general solutions taking into account the effect of 
replica symmetry breaking (RSB) in order to accurately assess 
the $L_p$-reconstruction. 

\section{Results}
\subsection{Theoretical predictions}
%\subsubsection{brief overview}
We numerically solved the saddle-point problem (\ref{freeenergy}) for various 
pairs of $\alpha$ and $\rho$ for $p=0,\, 1$ and $2$. 
When $\rho$ and $p$ are fixed, we always found only the successful solution 
$Q=m=\rho$ for sufficiently large $\alpha$. As we decreased $\alpha$, 
the successful solution lost its stability within the RS ansatz
and exhibited a transition to a failure solution. 

For the successful solution, conjugate variables $\widehat{Q}$ and $\widehat{m}$ are
always infinitely large. On the other hand, $\chi$ and $\widehat{\chi}$ do not necessarily 
diverge. We assessed the lower limit of the compression rate for given $\rho$, 
$\alpha_c(\rho)$,  for the $L_p$-reconstruction
by examining the local stability of the successful solution. 

\subsubsection{$p=0$}
The successful solution is stable if and only if $\alpha >\rho$. 
This indicates that $\alpha_c(\rho)=\rho$ for $p=0$. 
Even if the positions of the non-zero elements in $\bx^0$ are known, 
$\alpha > \rho$ is necessary for $\bF\bx=\by$ to have a unique solution. 
This implies that the $L_0$-reconstruction achieves
the best possible performance of compressed sensing.
However, the condition (\ref{AT}) always holds for the successful solution 
because of the discontinuous profile of $x_p^{*}(h;\widehat{Q})$.
This means that the assessment under the RS ansatz is 
not appropriate and further explorations taking RSB 
into account will be necessary before an
accurate performance evaluation of the $L_0$-reconstruction
can be made within our framework. However, this is beyond the scope of the current study.  
\subsubsection{$p=1$} $\widehat{\chi}$ of the 
successful solution is determined by 
\begin{eqnarray}
\widehat{\chi}&=&\alpha^{-1}
\left [
2(1-\rho)\left (
(\widehat{\chi}+1)\mathcal{Q}(\widehat{\chi}^{-1/2}) \right . \right . \cr
&\phantom{=}& \left . \left . 
-\widehat{\chi}^{1/2}\frac{e^{-1/(2 \widehat{\chi})}}{\sqrt{2\pi}}
\right )+\rho(\widehat{\chi}+1) 
\right ]
\label{determine_chi_hat}
\end{eqnarray}
where $\mathcal{Q}(x)=\int_x^{\infty}Dt$ is the conventional Q-function.  
Utilizing the solution of this equation, 
the stability condition of the successful solution is 
expressed as 
\begin{eqnarray}
\alpha > 2(1-\rho)\mathcal{Q}(\widehat{\chi}^{-1/2})+\rho
\label{L1capacity}
\end{eqnarray}
indicating that the reconstruction limit of 
the $L_1$-reconstruction is 
$\alpha_c(\rho)=2(1-\rho)\mathcal{Q}(\widehat{\chi}^{-1/2})+\rho$. 
$\alpha_c(\rho)$ also accords with the critical condition of 
(\ref{AT}), which implies that the RS solution is 
stable as long as (\ref{L1capacity}) holds. 
We thus conclude that there is no need of performing the 
RSB analysis for the performance evaluation of the $L_1$-reconstruction. 

We should emphasize that the reconstruction 
limit (\ref{L1capacity}) accords exactly with the limit
obtained in \cite{Donoho2006,DonohoTanner2009}
under the criterion of a successful reconstruction
that permits no reconstruction errors of any order by 
utilizing techniques of combinatorial geometry. 
This indicates that the reconstruction limit 
is not sensitive to the criterion of the 
successful reconstruction in the sense that 
it does not change even if we allow asymptotically negligible 
reconstruction errors as $P=\alpha N \to \infty$. 

Our results also imply that the replica method 
may be of utility not only to the current problem of 
compressed sensing but also to the whole field of 
combinatorial geometry. 

\subsubsection{$p=2$}
The successful solution is stable if and only if $\alpha \ge 1$, 
which indicates that $\alpha_c(\rho)=1$. 
For $\alpha > \alpha_c(\rho)=1$, (\ref{AT})
does not hold and, therefore, there is no need to do  
the RSB analysis. 
However, for $\alpha \ge 1$, one can 
perfectly reconstruct $\bx^0$ by using the 
Gauss elmination method to solve
the linear equation $\bF \bx=\by$. 
This leads us to conclude that the $L_2$-reconstruction has no 
capability of compressed sensing.

\begin{figure}[t]
\centerline{\includegraphics[width=7.5cm]{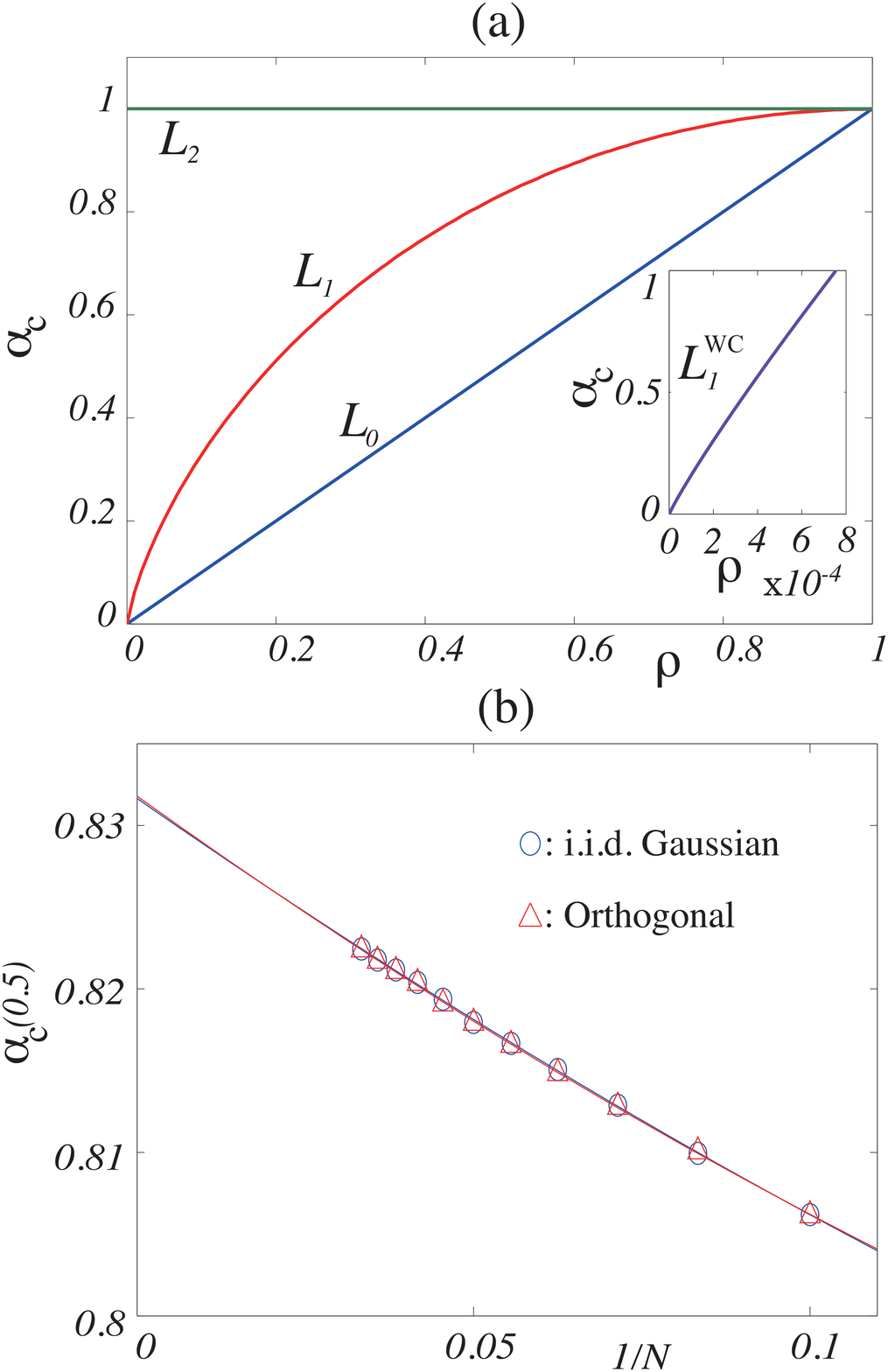}}
\caption{(a): Typical reconstruction limit for the $L_p$-reconstruction of $p=0,\,1$ and $2$. 
The inset represents the worst case upper-bound for the $L_1$-reconstruction offered by (\ref{sufficient1}) and (\ref{sufficient2}). 
(b): Experimental assessment of $\alpha_c(\rho=0.5)$
for the $L_1$-reconstruction
for (A): matrices of i.i.d. Gaussian random entries (circle) and 
(B): those of random orthogonal row vectors (triangle). 
A quadratic fitting with respect to $1/N$ yields
$\alpha_c(0.5)=\lim_{N \to \infty}\alpha_c(0.5,\,N)
\simeq 0.831\,65$ and $0.831\,79$ for (A) and (B), respectively. 
}
\label{fig2}
\end{figure}

%\vspace{1.5zh}

\subsection{Consideration}
Figure~\ref{fig2} (a) shows $\alpha_c(\rho)$ for 
$p=0,\,1$ and $2$. For comparison, 
the inset shows plots of (\ref{sufficient1}) 
and (\ref{sufficient2}) in the limit of $N,\,P \to \infty$ 
while maintaining $\alpha=P/N\sim O(1)$, $\rho=S/N \sim O(1)$.
In the case of large systems, 
the $L_1$-reconstruction is performable 
with a computational cost of $O(N^3)$ by utilizing interior point methods. 
On the other hand, although the $L_0$-reconstruction may potentially give a better reconstruction 
capability than the $L_1$-reconstruction, it is NP-hard. 
Figure~\ref{fig2} (a) and these indicate that 
the $L_1$-reconstruction is a practical method 
that possesses both computational feasibility
and relatively high reconstruction capability. 

There is a huge discrepancy between the typical reconstruction 
limit obtained here and the one for the worst case of \cite{CandesTao3,CandesTao2,Candes2008}, 
which is shown in the inset. This implies that there may be much room for improvement of the worst-case assessment, 
although we must keep in mind that the criterion of reconstruction success in the
current analysis, which permits asymptotically negligible reconstruction errors  as
$N \to \infty$, is different from that of the worst-case analysis in which no errors are allowed.

\subsection{Experimental validation}
{
To justify the result obtained above, 
we carried out numerical experiments 
on the $L_1$-reconstruction and the original 
vectors of $\rho=0.5$
for two matrix ensembles; 
(A): matrices of i.i.d. Gaussian random entries
and (B): those of random orthogonal row vectors.  
The results are summarized in Fig.~\ref{fig2} (b). 
%
%the results of which are summarized in Fig.~\ref{fig2} (b). 
%A trial of the experiments was performed as follows. 
%We generated an $N$-dimensional original vector 
%$\bx^0$ such that randomly chosen $S=\rho N$ elements were independently 
%sampled from a standard Gaussian distribution, 
%and the remaining $(1-\rho)N$ elements were set to zero. 
%We also prepared 
%an $N \times N$ matrix $\bF$ each entry of which 
%was an  i.i.d. Gaussian random variable of zero mean and variance $N^{-1}$. 
%
%a sample of $N \times N$ {\em random orthogonal matrix} as $\bF$ 
%for examining a representative case of $\bD=(\delta_{\mu i})$ in (\ref{SVD}), 
%where $\delta_{\mu i}=1$ for $\mu=i$ and $0$, otherwise. 
%%
%For numerically characterizing the criticality of the $L_1$-reconstruction, 
%the number of linear constraints of $\bF \bx= \by=\bF \bz^0$, $P$, was reduced one-by-one
%from $P=N$ and $P_c=P+1$ was recorded using the value of $P$ 
%at which $||\widehat{\bx}-\bx^0||_1 > 10^{-4}$ was satisfied first. 
%For searching $\widehat{\bx}$, we employed a package of 
%convex optimization schemes \texttt{CVX}
%\cite{CVX,DCP}. 
For both ensembles and each of $N=10,\,12,\,\ldots,\,30$, we numerically evaluated the critical 
compression rate for successful reconstruction for $\rho=0.5$,  
$\alpha_c(\rho=0.5,\,N)$, on the basis of $10^6$ experiments. 
We employed a package of convex optimization schemes \texttt{CVX} \cite{CVX,DCP}
for searching $\widehat{\bx}$. 
%The trials were repeated $10^6$ times for each $N$ 
%in order to assess experimental values 
%$\alpha_c(\rho=0.5,\,N)=\overline{P_c} /N$, 
%where $\overline{\cdots}$ denotes the arithmetic average over the trials. 
%The values for $N=10,\,12,\,\ldots,\,30$ were employed to estimate the reconstruction limit of $\rho=0.5$
%as $N \to \infty$. 
%The limit was determined by extrapolation of a quadratic fitting with respect to $N^{-1}$, 
The critical values of $N \to \infty$ were estimated by extrapolation of a quadratic fitting with 
respect to $N^{-1}$, which yielded $\alpha_c(0.5)=
\lim_{N \to \infty}\alpha_c(0.5,\,N)
\simeq 0.831\,65$ and $0.831\,79$ for (A) and (B), respectively. 
%This value is 
These values are in accordance with the theoretical prediction 
$\alpha_c(0.5)=0.831\,29\ldots$ up to the third digit (Fig.~\ref{fig2} (b)), 
which supports the universality of the critical relation. 
In \cite{CandesTao3}, 
the critical density $\rho_c$ for $\alpha=0.5$ is 
experimentally evaluated for 
relatively large systems of $N=512$ and $1024$
in the case of the i.i.d. Gaussian random entries.
}
Although the exact value is not provided, 
judging  by eye, 
the plots of the results are close to our theoretical prediction 
$\rho_c(\alpha=0.5)=0.192\,84\ldots$. 
This indicates that our approach at least has the ability to predict 
these experimental results with high accuracy. 
However, a general  
mathematical justification of the replica method 
is still to be done \cite{Talagrand}.

\section{Summary}
{
In summary, we assessed a typical reconstruction limit
of compressed sensing based on the $L_p$-norm 
reconstruction under linear constraints
for $p=0,\,1$ and $2$ in the limit of large systems
%. 
%For this, we used the replica method 
utilizing the replica method of statistical mechanics. 
%under the replica symmetric (RS) ansatz. 
%In our framework, the reconstruction limit is 
%evaluated by a local stability analysis 
%(within the RS ansatz) of 
%a solution that represents a successful reconstruction.   
For rotationally-invariant ensembles of the compression matrix $\bF$,  
our analysis indicates that the followings hold universally; the 
replica symmetric solution of the $L_0$-reconstruction 
achieves theoretically optimal reconstruction performance, 
but unfortunately, is unstable against perturbations that break 
the replica symmetry, Consequently, further analysis based on 
the replica symmetry breaking 
ansatz will be necessary for 
accurate assessment of its performance. 
The $L_2$-reconstruction has no capability of compressed sensing. 
On the other hand, the $L_1$-reconstruction has good reconstruction ability
which is characterized by a {\em universal} critical relation between 
$\alpha$ and $\rho$. 
Moreover, it is widely known that the $L_1$-reconstruction can be done 
at a computationally feasible cost.
These properties point to the great utility of the $L_1$-reconstruction in practice.

Our analysis supports that the universality of the critical relation
observed in \cite{DonohoTannerPhilTrans2009,DonohoTannerDCG2008}
holds for a certain class of the matrix ensembles irrelevantly to the details 
of the distribution of non-zero elements in the original vector $\bx^0$. 
However, the relations {\em can depend} on the details of $\bF$ when the 
ensemble is not rotationally invariant: this situation practically occurs when 
the original information to be compressed is expanded by non-orthogonal bases
and statistically independent sparse coefficients. 
An analysis on such cases will be reported elsewhere \cite{TakedaKabashimaISIT2010}. 
}
\section*{Acknowledgement}
This work was partially supported by Grants-in-Aid for Scientific Research on the
Priority Areas ``Deepening and Expansion of Statistical Mechanical Informatics'' 
from the Ministry of Education, Culture, Sports, Science and Technology, Japan
and KAKENHI No. 22300003 from JSPS. 

%%



%\bibliographystyle{IEEEtran}

\end{document}

%% file: commands.tex
\usepackage{amsfonts}

\newcommand\mbbm[1]{\mathchoice{\mbox{\boldmath{$\displaystyle #1$}}}%
		{\mbox{\boldmath{$\textstyle #1$}}}%
		{\mbox{\boldmath{$\scriptstyle #1$}}}%
		{\mbox{\boldmath{$\scriptscriptstyle #1$}}}}%
\newcommand{\bU}{\mbbm{U}}
\newcommand{\bV}{\mbbm{V}}

\newcommand{\bF}{\mbbm{F}}

\newcommand{\bD}{\mbbm{D}}

\newcommand{\bbs}{\mbbm{y}}

\newcommand{\bx}{\mbbm{x}}

\newcommand{\bbm}{\mbbm{m}}

\newcommand{\by}{\mbbm{y}}

\newcommand{\mR}{\mathbb{R}}
\newcommand{\mN}{\mathbb{N}}